# Casimir induced instabilities at metallic surfaces and interfaces

Kun Ding[1,2], Daigo Oue[1], C. T. Chan[2†], and J. B. Pendry[1*]

[1]The Blackett Laboratory, Department of Physics, Imperial College London, London SW7 2AZ, United Kingdom

[2]Department of Physics, Hong Kong University of Science and Technology and Center for Metamaterials Research, Clear Water Bay, Kowloon, Hong Kong, China

†Email: phchan@ust.hk

*Email: j.pendry@imperial.ac.uk

**Abstract**

Surface plasmons subject to a surface distortion split asymmetrically in energy resulting in a net lowering of zero-point energy. This is because surface plasmon eigenvalues are the square of frequencies, a statement generally true for electromagnetic excitations. We utilize the method based on conformal mapping to demonstrate asymmetric splitting under surface corrugations leading to a decrease in zero-point energy of a single corrugated metallic surface contributing to surface reconstructions but too small on its own to drive the reconstruction. However, by introducing a second metallic surface more significant lowering of energy is seen sufficient to drive the instability of a mercury thin film.

*Introduction.* Casimir forces [1,2] arise from zero-point quantum fluctuations of the electromagnetic field and play a central role in biology, adhesion, friction, wetting and a host of other phenomena. Here we are concerned with the impact of Casimir forces on the stability of surfaces in the presence of distortions of various kinds. We are particularly interested in the role of surface plasmons. We contrast the Casimir bonding mechanism with electronic bond formation. When two hydrogen atoms meet and interact their energy levels split: $E \pm D$. The antisymmetric mode is raised in energy and the symmetric mode depressed. To a first approximation the shifts in energy are of equal but of opposite sign. A populated lower and unpopulated upper level are responsible for the bonding energy. In contrast two helium atoms do not form an electronic bond as the Fermi exclusion principle requires all four electrons concerned to populate both lower and upper levels and hence no bonding. However, two helium atoms do attract via the Casimir force. It is instructive to ask why this is so. Helium atoms have a dipole polarizability and the dipole modes of two atoms split as they approach just as electronic energy levels do, but with a key difference. In the equations governing electromagnetic fields the frequency appears as $w^2$ and it is this quantity which is approximately equally split by the interaction of levels: $w^2 \pm D$ so that even though each level is equally populated as in the electronic case, there is a net lowering of energy of $-\hbar\Delta^2/8\omega^3$ as a consequence of taking a square root. This argument applies to any splitting of electromagnetic levels and accounts for the almost universal attractive nature of dispersion forces.

With this argument in mind, it is to be expected that when a surface is distorted the ensuing splitting of electromagnetic levels will favor the distortion and the ensuing instability. Of course, other often more powerful forces may oppose the distortion but as we shall show the Casimir instability sometimes wins. Such competition is experimentally observable because the recent development of nanofabrication and measurement technique already bring some elaborate designs of fluctuation-related phenomena into reality [3-7], such as Casimir torque [8], non-monotonic Casimir force in gratings [9], quantum trapping [10]. These works reveal that geometric modifications together with functional materials offer an excellent platform to manipulate the Casimir force. However, the theoretical study of corrugation effects requires a reliable and fast method to calculate the electromagnetic scattering matrix of corrugated surfaces because many repeated calculations are required [13-15]. Owing to its computational speed and accuracy, we employ a method based on transformation optics in its conformal mapping realization to calculate the scattering matrix of corrugated surfaces and investigate the role of asymmetric splits in the Casimir energy. We demonstrate that the splitting of plasmonic modes under corrugation indeed decreases surface energy of a single surface. To make such effects experimentally observable, we introduce a second surface, and demonstrate that the asymmetric split can cause the decrease of Casimir energy for the gold cavity system and create instabilities in a mercury thin film.

*Asymmetric splits of plasmonic modes.* We first introduce a conformal mapping technique that can calculate the scattering matrix of corrugated surfaces very efficiently [16]. We start from a single corrugated surface with its profile given by a conformal mapping of the form

$$z = \Gamma \ln \frac{1}{e^w - iw_0}, \tag{1}$$

which transforms a flat surface (with the air/metal boundary at $u = u_0$ in the slab frame as shown in Fig. 1a) to a corrugated surface. Here, $w = u_0 + iv$ and $z = x + iy$ are the surface coordinates in the slab and metasurface frame, respectively. As shown in Fig. 1(a), the parameter $\Gamma$ determines the corrugation period $a = 2\pi\Gamma$, and the modulation strength $A$, defined as the distance between the top of the protrusion and the average position, is proportional to $w_0$ ($|w_0| \leq e^{u_0}$). When $w_0$ is small, the profile of the corrugated surface is almost a sinusoidal function with $A = \Gamma w_0/e^{u_0}$. We use the Drude model to describe the metal throughout this work.

To confirm the aforementioned asymmetric split of plasmonic modes, we calculate the eigenfrequencies of the plasmonic modes for the corrugated surface by finding the zeros of $\det(\mathbf{R}_1^{-1})$ in the real frequency axis for a given wavevector. Here, $\mathbf{R}_1$ is the reflection matrix of the surface in Fig. 1(a) using a Fourier representation. The dimension of $\mathbf{R}_1$ is $2(2n_{cg} + 1)$, where $n_{cg}$ is the cutoff of the basis of Bloch wavenumber $G = 2\pi/a$, and the factor 2 outside the bracket accounts for two polarizations. Low-index gold surfaces reconstruct to rather complex surface unit cells [17-20], and in order to investigate whether the asymmetric splits of plasmonic modes play a role in the reconstruction, we choose our geometric parameters to mimic the surface profile of the reconstructed Au(100) surface with a corrugation period $a = 14.4$Å and modulation strength $A = 1.44$Å [17]. Figure 1(b–c) shows the plasmonic dispersion of such a corrugated plasmonic surface (subtracting the surface plasmon frequency $\omega_{sp}$) and we note that the corrugation splits the plasmonic modes to lower (red squares) and upper (blue circles) plasmonic branches. The sum of their frequencies is shown by magenta dots in Fig. 1(b–c). We see that in the $k_z\Gamma = 0.0$ plane, the downshift in frequency of the lower branch is larger than the upshift of the upper branch, indicating that their sum is negative with respect to $2\omega_{sp}$. This confirms the aforementioned argument that the square root of $\omega^2$ makes the sum of $\omega$ negative, as shown by magenta dots in Fig. 1(b). We see that the asymmetric split increases when $k_y$ approaches the Brillouin zone (BZ) boundary. Along the homogeneous z-direction, $k_z$ increases without bound in a continuum model. However, the corrugation-induced split becomes almost symmetric when $k_z$ increases to a large value, as shown in Fig. 1(c). Two plasmonic modes and the sum of their frequencies as a function of $k_z$ for $k_y\Gamma = 0.5$ are shown in Fig. 1(c). We note that $k_z$ is normalized by the Fermi wavenumber $k_F$, and its value is $k_F\Gamma = 2.78$. We see that the asymmetric split approaches zero when $k_z > 0.3k_F$, indicating that the corrugation induced splitting effect rapidly approaches zero when $k_z$ is large.

***Plasmonic contributions to reconstructed gold surfaces.*** The difference in plasmonic zero-point energies should be taken into account when we consider the change of the ground state surface energy during a surface reconstruction process. We see that the corrugation induced asymmetric splits of plasmonic modes are significant in a certain region of the BZ, and in order to calculate such plasmonic mode contributions to the zero-point energy, we count the modes using the following expression

$$E = \frac{1}{(2\pi)^2}\int_{-\pi/a}^{+\pi/a} dk_y \int_{-k_{zc}}^{+k_{zc}} dk_z \frac{\hbar}{2\pi}\int_0^{\omega_p} d\omega\, \mathrm{K}_{sg}(\omega, k_y, k_z), \qquad (2)$$

$$\mathrm{K}_{sg}(\omega, k_y, k_z) = \sum_i \mathrm{Im}\left[\ln(\lambda^i)\right], \qquad (3)$$

where $\lambda^i$ are eigenvalues of reflection matrix $\mathbf{R}_1$ and $k_{zc}$ is the chosen cutoff value of $k_z$. Equation (3) and $\det(\mathbf{R}_1^{-1})$ do essentially the same task of capturing the plasmonic modes. We plot in Fig. 2(a) the calculated $\mathrm{K}_{sg}$ as a function of $\omega$ for the flat (dashed lines) and corrugated gold (solid lines) surfaces. When the loss is small, *i.e.* $\gamma$=1e-4 meV, the $\mathrm{K}_{sg}$ for the flat surface is a step function, with the sharp change occurring at $\omega_{sp}$, as shown by the gray dashed line in Fig. 2(a). Corrugations induce two additional step-like features (solid gray line in Fig. 2a), indicating the emergence of two plasmonic modes in accordance with the results in Fig. 1. Increasing $\gamma$ to 35 meV, the step edges are smoothed, but the salient features of plasmonic modes remain, as shown by the blue lines in Fig. 2(a). This confirms that the kernel function defined in Eq. (3) successfully captures the plasmonic modes of these surfaces.

We then perform the integral (2) with Eq. (3) to investigate sum of plasmonic mode frequencies up to the bulk plasmon frequency $\omega_p$. As a control calculation, we confirm that the kernel function and integral method successfully reproduces the essential physics for a flat gold surface (See Section I in Supplemental Material [21]). We then apply Eq. (2) to investigate the zero-point energy plasmonic contributions of a corrugated surface. Figure 2(b) shows the corrugation induced change of zero-point energy $\Delta E\ (= E(\mathrm{w}_0) - E_0)$ for different values of $k_{zc}$. For a fixed cutoff $k_{zc}$, the decrease of zero-point energy converges to a finite number with the increase of $n_{cg}$, indicating that the energy differences between flat and corrugated surfaces approach a converged result for a reasonable number of Fourier components along the corrugation (y)-direction, and then the only physically meaningful cutoff is $k_{zc}$. The asymmetric split of plasmonic modes shown in Fig. 1 suggests that the decrease of surface energy should approach a finite number when $k_z$ is large enough, which is confirmed in Fig. 2(b). We see that the change of zero-point energy converges to $-0.2 \, \text{meV}/\text{Å}^2$ when $k_{zc} > 0.2 k_\text{F}$, confirming that the splitting of plasmonic modes under geometric corrugations decreases the zero-point energy. We note that there is always a momentum cutoff above which we need to go beyond a continuum model, and we need to solve the plasmon problem using quantum many-body formulations. For Au, that cutoff is ~ $0.6 k_\text{F}$ (see Section II in Supplemental Material [21]), and the classical approach here converges (at $k_{zc} \sim 0.2 k_\text{F}$) well before this limit.

*Casimir energy of a metallic cavity.* We see that the zero-point energy of a corrugated surface is lower than that of flat surface due to the asymmetric splitting of plasmonic modes. However, the decrease is small compared with the typical surface energy change in a reconstruction process and hence the zero-point energy of a single air/metal boundary cannot drive metal surface reconstruction. We will show below that Casimir energy change can induce surface corrugation in a metal cavity system defined in Fig. 3(a). The distance $L$ between two surfaces is defined based on the average position. Before performing rigorous calculations, we first estimate the role of corrugations using the commonly-used proximity force approximation (PFA) [14,22]. PFA calculates the Casimir energy by discretizing the corrugated surface into several small flat sectors and add the plane-plane Casimir energy contributions together to give the total energy. The fact that the Casimir energy between two perfect mirrors is proportional to $-L^{-3}$ can give us an intuitive understanding of the corrugation effect. Since the dependence of Casimir energy is not linear in $L$, the protruded portion lowers the Casimir energy more than the increase of energy of the concave region, giving an overall decrease of Casimir energy when corrugations are present. To characterize this, we define the quantity $C_w$ as $E(w_0) = E_0 + C_w A^2$, where $E_0$ is the Casimir energy for the cavity containing two flat surfaces ($w_0 = 0$) and $E(w_0)$ is that of the cavity with one corrugated surface ($w_0 \neq 0$) with $L$ fixed. The calculated $C_w$ by PFA as a function of $a$ for $L = 15$ nm and 25 nm are shown by filled and open stars in Fig. 3(b). We see that $C_w$ is negative for all the periods, and the magnitude of $C_w$ for $L = 15$ nm is larger than those for $L = 25$ nm, consistent with the intuitive understanding. We also note that $C_w$ does not change much when varying $a$ owing to the way PFA handles the problem.

We then calculate the corrugation induced change in Casimir energy rigorously using the conformal mapping technique. The Casimir energy of such a cavity system can be calculated using the Lifshitz's mode counting function $K_{cv}(\omega = i\xi, k_y, k_z) = \ln \det[\mathbf{I} - \mathbf{R}_1 \mathbf{R}_2]$ in the integral (2) without any cut-offs in frequency and wavenumbers, where $\mathbf{R}_1$ and $\mathbf{R}_2$ are reflection matrices of surface $O_1$ and $O_2$ using the Fourier basis. In our formulation, the propagation phase matrix is already embedded implicitly in $\mathbf{R}_2$, and the integral is performed as usual in the imaginary frequency axis. The calculated $C_w$ for $L = 15$ nm and 25 nm are shown by filled and open red dots in Fig. 3(b). When $a$ is large, the PFA results agree well with the rigorous Lifshitz result. This shows that PFA can give reasonably reliable Casimir energies when the modulation period is in the micron scale. However, corrugation always increases the surface area, and the surface energy due to bond breaking must increase correspondingly. Whether a metal surface can lower its surface energy by corrugation is then a competition between the decrease due to Casimir energy (Fig. 3b) and the increase due to electronic surface energy. We can express the change of energy to the leading order as $\Delta E = (C_w + C_{sf})A^2$, where $C_w$ and $C_{sf}$ are contributions from Casimir energy and intrinsic surface energy, respectively. The intrinsic surface energy contribution $C_{sf}$ is proportional to the increase of surface area and equal to $2\pi^2 \gamma_{sf}/a^2$, where $\gamma_{sf}$ is the surface tension coefficient. $C_{sf}$ is positive and decays parabolically in $a$, while the Casimir energy contribution $C_w$ is almost a negative constant for $a$ in the micron scale. Therefore, the surface will be corrugated if another surface $O_2$ is placed close enough to the surface.

It is important to note that the values of $C_w$ calculated by the Lifshitz formula are much more negative than those by PFA when the modulation period is less than 100nm, indicating that corrugations induce a Casimir energy decrease more rapidly than pair-wise additive estimations like the PFA. This shows that the surface will be MORE corrugated than the results predicted by PFA because of the additional lowering in energy that is computed accurately by our method. Such a disagreement is partly because PFA ignores the corrugation-induced asymmetric splits in plasmonic modes of a single surface. Figure 3 shows the zero-point energy lowering due to the coupling with another surface, while the result in Fig. 2 is an intrinsic single boundary lowering in zero-point energy. Both of them share the same mechanism if we understand them as the $\omega^2$ eigenvalue perturbation problem.

*Instabilities of a mercury film.* We now consider another example of a pair of metal/dielectric boundaries. We consider the stability of a mercury film with a thickness $d$, embedded between two dielectric media $\varepsilon_1$ and $\varepsilon_2$ (see inset of Fig. 4). We assume that the upper surface has a corrugated profile $A\sin(2\pi y/a)$, and again express the change of energy as $\Delta E = (C_w + C_{sf})A^2$, where $C_w$ here is the contribution from the Casimir energy change due to the coupling of the plasmons at the two metal/dielectric interfaces. As $C_{sf}$ is positive and decays parabolically in $a$, while $C_w$ is almost a negative constant for the micron size $a$, a mercury film with a given thickness will become unstable because the Casimir energy will dominate when the corrugation period $a$ is large than a critical value. To quantify the instability, we calculate $C_w$ within PFA together with quasistatic approximation and then plot the $C_w + C_{sf} = 0$ line by the solid blue line in Fig. 4 for $\varepsilon_1 = \varepsilon_2 = 1$, which confirms the above statements. The mercury film is usually placed on some substrate, so we also plot the $C_w + C_{sf} = 0$ line for $\varepsilon_2 = 2$ and 3 by the solid red and magenta line, respectively, while keeping $\varepsilon_1 = 1$. Compared with the $\varepsilon_2 = 1$ case, we see that the critical values of $a$ increase as $\varepsilon_2$ increases. All these results show that a flat thin mercury film is unstable against corrugation in the large scale. This is an example of a Rayleigh-Taylor instability [24,25].

*Conclusions.* We demonstrated that surface corrugations induce plasmonic modes of metal surfaces to split unevenly because the equation of motion of plasmons is second order in time which is generally true for electromagnetic excitations. Such an asymmetric split always decreases the zero-point energy of a single air/metal interface and can give rises to the decrease of the Casimir energy for a cavity system consisting of two metallic surfaces. It also contributes to the instability of a mercury thin film against corrugation. Since such an asymmetric mode splitting is intrinsic to the coupling of plasmonic modes, their contributions can be observed in other hybridized plasmonic systems, leading to one kind of electromagnetic bonding, and can cause the attraction of neutral metallic objects. Last but not least, accurate calculations of zero-point and Casimir energies of complex interfaces usually require a formidable amount of computer resources, and our calculations are made tractable by the conformal mapping technique which can provide an excellent platform to investigate other electromagnetic fluctuation-type problems.

**Acknowledgements**

K.D. and J.B.P. acknowledge support from the Gordon and Betty Moore Foundation. D.O. acknowledges support from the Imperial College President's Scholarship. K.D. and C.T.C. acknowledge funding from Research Grants Council (RGC) Hong Kong through grants AoE/P-02/12 and 16303119.

**References**

[1] J. N. Israelachvili, Intermolecular and Surface Forces, (Academic, San Diego, 1998).

[2] A. W. Rodriguez, F. Capasso, and S. G. Johnson, Nature Photonics **5**, 211 (2011).

[3] S. K. Lamoreaux, Phys. Rev. Lett. **78**, 5–8 (1997).

[4] H. B. Chan *et al.*, Phys. Rev. Lett. **101**, 030401 (2008).

[5] A. Lambrecht and V. N. Marachevsky, Phys. Rev. Lett. **101**, 160403 (2008).

[6] J. N. Munday, F. Capasso, and V. A. Parsegian, Nature **457**, 170–173 (2009).

[7] A. A. Banishev, J. Wagner, T. Emig, R. Zandi, and U. Mohideen, Phys. Rev. Lett. **110**, 250403 (2013).

[8] David A. T. Somers, Joseph L. Garrett, Kevin J. Palm, and Jeremy N. Munday, Nature 564, **386** (2018).

[9] L. Tang, M. Wang, C. Y. Ng, M. Nikolic, C. T. Chan, A. W. Rodriguez, and H. B. Chan, Nat. Photonics **11**, 97 (2017).

[10] Rongkuo Zhao, Lin Li, Sui Yang, Wei Bao, Yang Xia, Paul Ashby, Yuan Wang, and Xiang Zhang, Science **364**, 984 (2019).

[11] F. Intravaia, C. Henkel, and A. Lambrecht, Phys. Rev. A **76**, 033820 (2007).

[12] P. Nordlander, C. Oubre, E. Prodan, K. Li, and M. I. Stockman, Nano Letters **4**, 899 (2004).

[13] L. M. Woods *et al.*, Rev. Mod. Phys. **88**, 045003 (2016).

[14] Astrid Lambrecht, Paulo A. Maia Neto, and Serge Reynaud, New J. Phys. **8**, 243 (2006).

[15] M. T. H. Reid, A. W. Rodriguez, J. White, and S. G. Johnson, Phys. Rev. Lett. **103**, 040401 (2009).

[16] J. B. Pendry, Paloma A. Huidobro, and Kun Ding, Phys. Rev. B **99**, 085408 (2019).

[17] Xiao-Qian Wang, Phys. Rev. Lett. **67**, 3547 (1991).

[18] J. V. Barth, H. Brune, G. Ertl, and R. J. Behm, Phys. Rev. B **42**, 9307 (1990).

[19] Ch. Wöll, S. Chiang, R. J. Wilson, and P. H. Lippel, Phys. Rev. B **39**, 7988 (1989).

[20] U. Harten, A. M. Lahee, J. Peter Toennies, and Ch. Wöll, Phys. Rev. Lett. **54**, 2619 (1985).

[21] See Supplemental Material at *link* for numerical details.

[22] B. V. Derjaguin, Untersuchungen über die reibung und adhäsion. Kolloid Z. **69**, 155–164 (1934).

[23] Bong-Ok Kim, Geunseop Lee, E. W. Plummer, P. A. Dowben, and A. Liebsch, Phys. Rev. B **52**, 6057 (1995).

[24] J.W.S. Rayleigh, Proceedings of the London Mathematical Society, **14**, 170–177 (1883).

[25] G.I Taylor, Proceedings of the Royal Society of London. Series A, Mathematical and Physical Sciences. **201** (1065) 192–196 (1950).

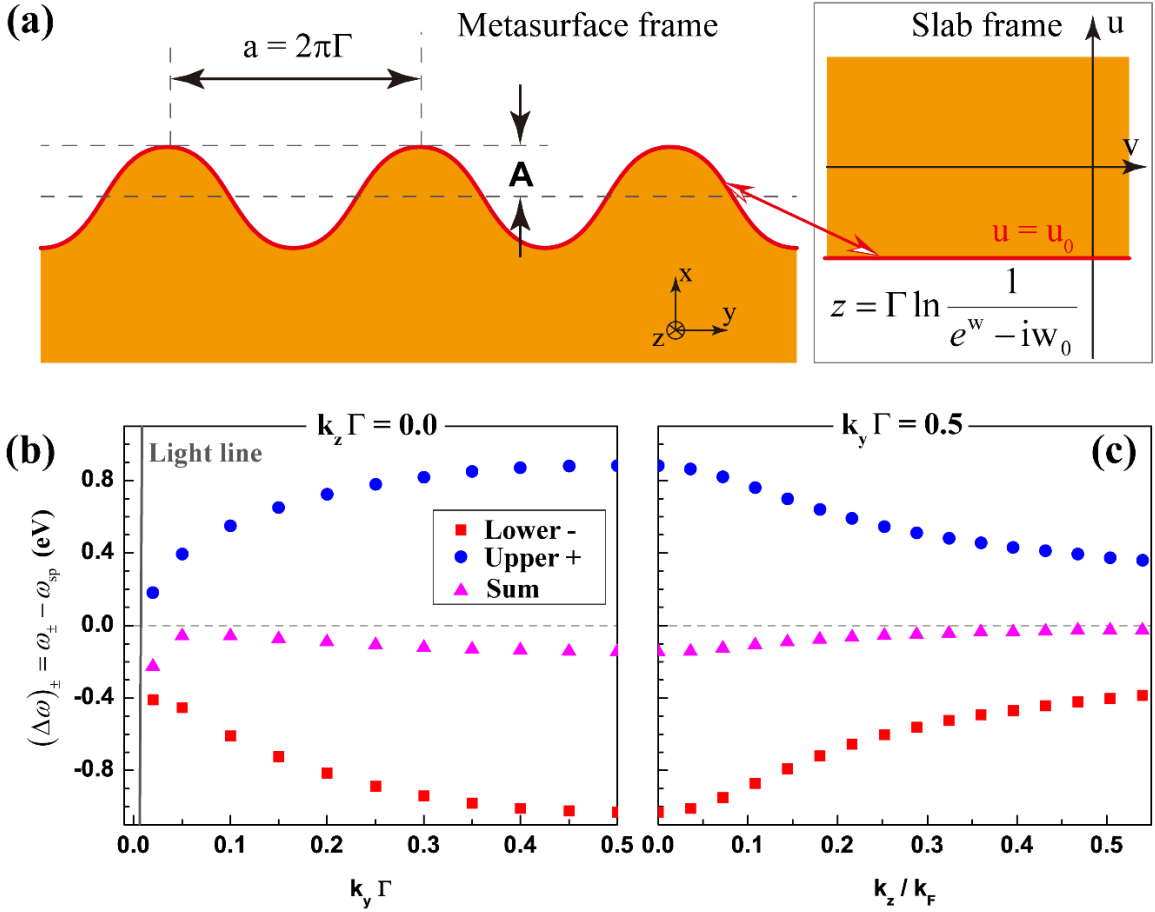

Figure 1. (a) The schematic picture of a corrugated surface. Its profile in the metasurface frame is generated from the slab frame using a conformal mapping. The corrugation period $a$ and modulation strength $A$ of the corrugated surface are defined in the figure. (b–c) The eigenfrequencies of two plasmonic modes for a corrugated surface with respect to $\omega_{sp}$ in the $k_z\Gamma = 0.0$ plane (b) and $k_y\Gamma = 0.5$ plane (c) are plotted by filled squares and circles. The magenta triangles show the sum of these two modes. The geometric parameters are $\Gamma = 2.291\text{Å}$, $w_0 = 0.557$, and $u_0 = 0$. We use a Drude model with $\omega_p = 9.0$ eV and $\gamma$=1e-4 meV.

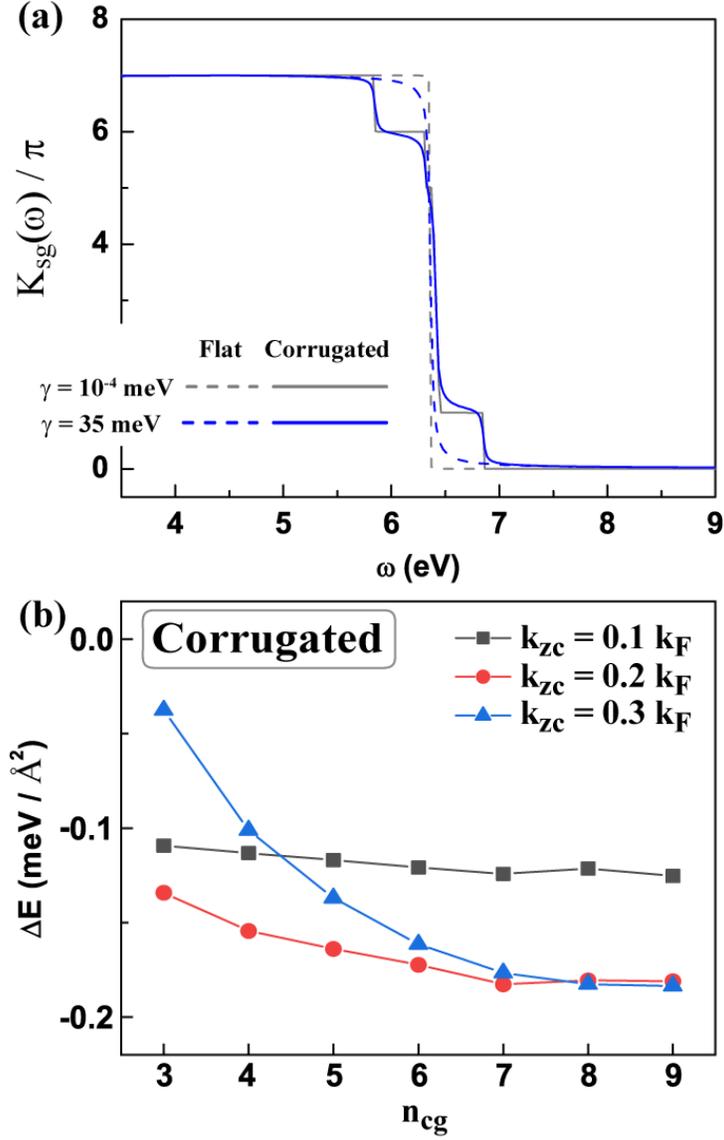

Figure 2. (a) The kernel function $K_{sg}$ as a function of frequency for $k_y\Gamma = 0.1$ and $k_z\Gamma = -0.8$. The cutoff in Fourier order is $n_{cg} = 3$. The gray and blue lines correspond to γ=1e-4 meV and γ=35 meV, respectively. (b) The decreases of surface energy for three different cutoff values of $k_{zc}$ are shown by solid gray, red, and blue lines.

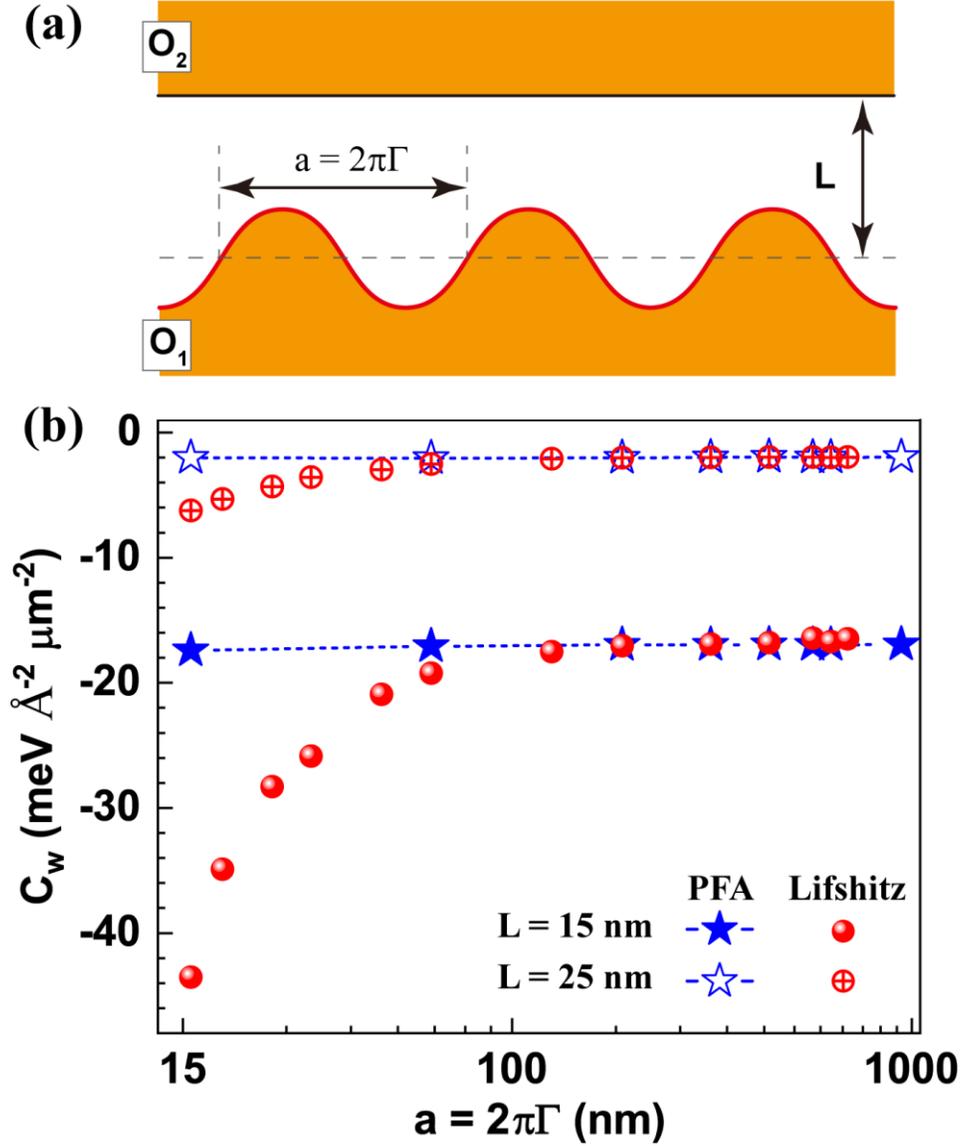

Figure 3. (a) The schematic picture of the metallic cavity with one corrugated surface and one flat surface. (b) The coefficients $C_w$ as a function of the corrugation period $a$ calculated by PFA and Lifshitz formula are shown by blue stars and red circles, respectively. The filled and open markers are results for $L$ = 15 nm and 25 nm, respectively.

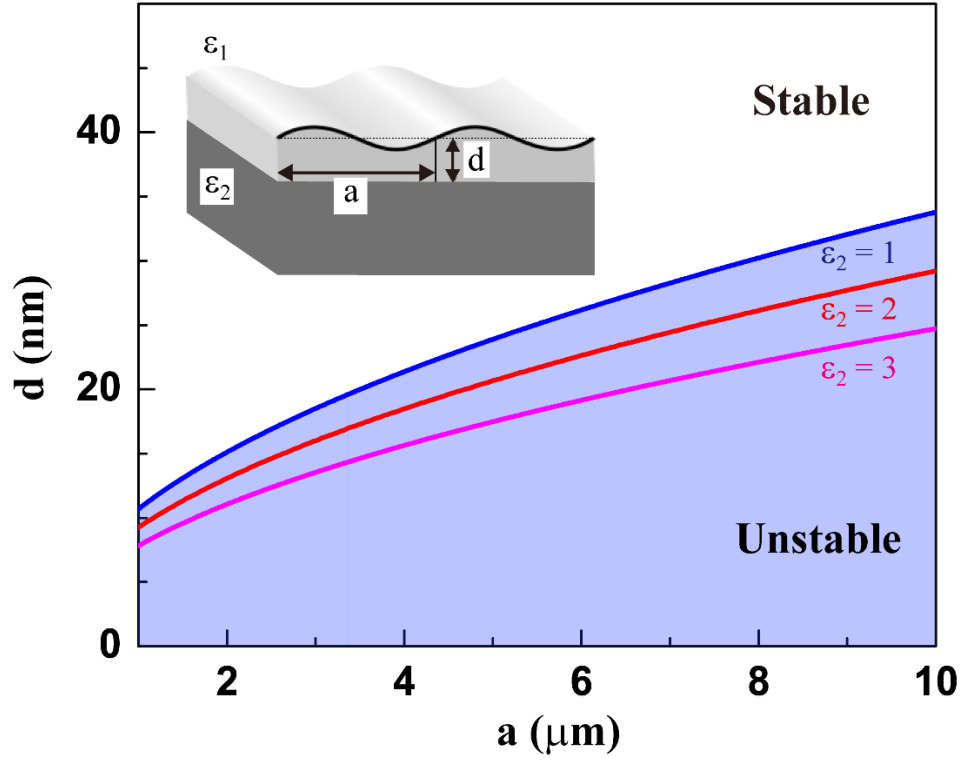

Figure 4. The $C_w + C_{sf} = 0$ line in the thickness $d$ and period $a$ plane for $\varepsilon_2 = 1$, 2, and 3 are shown by solid blue, red, and magenta lines, respectively. The inset is the schematic picture of the thin film embedded between two dielectric media $\varepsilon_1$ and $\varepsilon_2$. The parameters of mercury used are $\omega_{sp} = 6.83$ eV, and surface tension coefficient $\gamma_{sf} = 27.6$ meV/Å$^2$ taken from Ref. [23].